\documentstyle[aps,prl,multicol,epsf]{revtex} 
\def\frac#1#2{{\textstyle{#1\over#2}}} 

\title{ \rightline{ \small DAMTP-2000-12 \normalsize} 
\rightline{\small hep-ph/0009130 \normalsize} 
\centerline{Problems with empirical bounds for strangelet production
at RHIC} } 

\author{ Adrian Kent} 
\address{ 
Department of Applied Mathematics and 
Theoretical Physics, University of Cambridge,\\ 
Silver Street, Cambridge CB3 9EW, U.K.\\ 
} 

\date{September 10, 2000} 
\begin{document} 
\maketitle 
\begin{abstract}
Recent papers by Busza et al. and Dar et al. 
have considered empirical bounds on the risk of producing
a hypothetical stable negatively charged strangelet at
the Brookhaven relativistic heavy ion collider (RHIC) experiments, 
and thereby destroying the Earth.  
We examine here ways in which these bounds could, 
hypothetically, be evaded.  We find several, some of which 
have not previously been considered.  
These possible flaws do not affect the theoretical arguments against the 
existence of negatively charged strangelets or the possibility
of producing them at RHIC if they were to exist.  
\end{abstract} 
\vskip10pt 
\section{Introduction}\label{introduction} 

\subsection{Background}

High energy collider experiments tend to explore new 
realms of physics, in which theory tends to be uncertain.  
Catastrophic phenomena clearly occur in nature, and
with sufficient speculation one can imagine theoretical
mechanisms by which catastrophe could ensue from a collider experiment.
Such mechanisms, of course, tend to be discussed and investigated
for plausibility whenever any new experiment is proposed. 

The experiments considered here are relativistic heavy ion collisions
scheduled to take place at Brookhaven (RHIC) and CERN (ALICE).  
Several speculative catastrophe mechanisms have been discussed.
We examine here the one which seems at first sight the least 
implausible and which has attracted most attention, namely
the possibility that, if negatively charged metastable 
strange matter exists, it could be produced in the 
experiments and go on to destroy the Earth in a runaway reaction. 

This possibility was investigated by 
Busza et al. (BJSW), who were requested to review candidate mechanisms for
catastrophe scenarios at RHIC following public concern.  
Their initial report\cite{bjswone} concluded
that the possibility of strangelet-induced catastrophe can be
firmly excluded by compelling theoretical arguments, and that
very large safety factors can also be derived from the survival
of the Moon and other astrophysical evidence.
This analysis partly relies on independent work by Dar et
al.\cite{ddh} (DDH), who noted a theoretical hypothesis
which would imply that the survival of the Moon would provide 
no useful risk bounds, but concluded that the observed rate
of supernovae provides large safety factors even under this
hypothesis.

Following criticisms, Busza et al. produced 
a revised version\cite{bjswtwo} of their report. 
In this version, BJSW shift their ground considerably, explicitly 
disclaiming any attempt to decide what might be an acceptable 
bound on the risk of 
catastrophe, and acknowledging explicitly the possibility 
that all empirical bounds on that risk would be 
invalidated if the hypothetical strangelets had particular
unfortunate properties.  However, they underline their view 
that compelling theoretical arguments provide sufficient, 
albeit unquantifiable, reassurance that there will be no catastrophe.  

\subsection{Scope of this paper} 

Considering the possibility, however remote, of so large a catastrophe
is particularly difficult, since normal scientific practice is in
some important respects inadequate.  In everyday research, a hypothesis which
has perhaps at best a $1 \%$ chance of being correct is often
dismissed as not worth pursuing, and a scenario which relies on two
or three such hypotheses is very unlikely to be taken seriously.  
Such judgements have to be made: science would progress 
much more slowly if disproportionate attention was paid to 
unlikely hypotheses and implausible explanations.   
However, when trying to exclude the possibility of 
global catastrophe, a $10^{-2}$, $10^{-4}$ or even $10^{-6}$ 
probability of error is far from negligible, indeed alarmingly high.  
So every argument and every piece of conventional wisdom
needs to be very carefully scrutinised in case 
it might have some neglected flaw.  

A good way of trying to uncover such flaws, which I gather is
often used in analysing risks of large disasters, is to 
give different roles to different teams of people.\cite{calogero}
One team is required to make the case for safety.
The other is required to try to find flaws in that case, 
to look out for the unexpected, to search for ways in which
things could go wrong --- in short, to play devil's advocate. 
This paper is written in the spirit of a contribution from the
second, contrarian, team.  

The discussion here focusses on the empirically based arguments for 
bounds on the risk of catastrophe, which 
form an important part of the case made for safety at RHIC and have 
been widely publicized in an attempt to allay public fears.  
Its aim is to probe the robustness of those arguments, not 
to argue that the risk is in fact overly large.  

I also consider a potential danger that could arise from
the production of positively charged strangelets.  
While the risk of catastrophe arising this way appears pretty small
--- the strangelets would have to be somehow transported to the
Sun, or introduced into an artificial fusion reaction --- it 
deserves careful consideration.

This paper does not examine whether
the reassurance given by theoretical arguments is sufficient 
in itself, as BJSW suggest.\cite{bjswone,bjswtwo}
Nor does it consider what might be an acceptable bound on 
the risk of catastrophe.  
These questions will be addressed 
elsewhere.\cite{akelsewhere}

\subsection{Empirically derived risk bounds: general comments} 

Approximately $2 \times 10^{11}$ high energy collisions of pairs
of gold ions are scheduled to take place over the $10$ year
lifetime of RHIC.  Following BJSW and DDH,\cite{bjswone,ddh,bjswtwo} 
we consider the speculative ``killer strangelet'' 
scenario in which the experiment might lead to disaster. 

The scenario requires the following hypotheses.  Small atomic mass metastable 
negatively charged strangelets must exist.  It must be
possible to produce them in heavy ion collisions.  They must 
be sufficiently stable to survive long enough after production
to interact with matter in the surrounding environment.  
Then, strange matter must be stable in bulk, with a spectrum
of stable strangelets such that a negatively charged metastable 
strangelet can begin a runaway reaction by fusing successively with
positively charged nuclei, producing larger negatively charged
strangelets, and eventually consuming the Earth.  

Empirical bounds on the risk of such a catastrophe can be
derived because the type of collisions planned at RHIC also
occur naturally when high energy cosmic rays collide, either
with each other or with heavy nuclei in celestial bodies.
If the probability of dangerous strangelet production in
such collisions is sufficiently large, natural catastrophes 
should be induced, and we should be able to observe the 
consequences.  The fact that we see no evidence of 
strangelet-induced catastrophes means we can
bound the probability of strangelet production per collision,
and hence the overall probability of catastrophe at RHIC.
As quantum processes are probabilistic, 
small bounds on the probability of catastrophe are
the best we can hope for from this line of reasoning.
Empirical bounds cannot show that the probability 
is zero. 

Moreover, no bound can be established
from empirical data alone.  Any argument establishing a 
bound must also rely on some theoretical hypotheses.
Hence, to establish a bound of the form $p_{\rm catastrophe} < p_0$, 
we need to have at least the same level of confidence in all 
the relevant hypotheses.  It will immediately be seen that 
this is likely to pose a serious problem, since on the one hand
no bound will be thought reassuring unless $p_0$ is very small indeed,
and on the other the necessary hypotheses involve a physical regime 
where theory is fairly uncertain.  

The failure to consider
this problem carefully and systematically is one of the
weakest and most troubling points of BJSW and DDH's analyses of the
empirical evidence.  The discussion below highlights some gaps 
in their arguments.   

\section{Risk bounds derived from the survival of the Moon} 

BJSW first consider the value of the Moon's survival as empirical
evidence against the possibility of catastrophe at RHIC.  
They note that cosmic rays, including high energy heavy 
nuclei, are constantly bombarding the Moon.  The cosmic ray flux
is believed to have been at the present level or higher during the
Moon's lifetime, roughly $5$ billion years.  Since the Moon, 
unlike the larger planets in the solar system, is unprotected
by an atmosphere, these cosmic rays collide with 
its surface, which contains a significant component
of heavy nuclei and, in particular, is iron-rich.  
To derive a risk bound from these observations requires a series 
of assumptions.  

\begin{itemize}

\item We need to decide precisely which types of 
heavy nuclei collisions should be considered good models for
those at RHIC (or other collider experiments).  RHIC involves
gold-gold collisions.  BJSW suggest considering all nuclei with
$Z > 70$ as `gold', that is, effectively equivalent to gold
for the purposes of this analysis.  They also suggest, as a less
conservative alternative, taking iron (Z=26) as effectively equivalent to
gold.  

BJSW then produce estimates the rate of relevant collisions on the
Moon by the following reasoning. 
The flux of iron nuclei in cosmic rays at the relevant energies
is fairly reliably known, and the flux of `gold' nuclei can 
be estimated by extrapolating from lower energy data.  
The iron abundance on the Moon can conservatively be estimated by 
taking the lowest value of the abundance measured at six Apollo
spacecraft landing sites; BJSW round this value down 
from 4.2\% to 4\%. 

Accepting BJSW's estimates and extrapolations, and the equivalence 
of `gold' to gold, then gives four possible bounds, 
derived by considering respectively iron-iron, iron-`gold', `gold'-iron
and `gold'-`gold' collisions.  
BJSW consider the first and third of these.  
The fourth, which most closely parallels the RHIC
experiments, turns out to give no useful bound at all.
BJSW note this important caveat in their analysis, but 
neglect it when summarising their conclusions.  

\item  Any strangelets produced with high speed in 
Moon centre-of-mass frame may break up while being slowed down
by lunar matter; if so, they will not survive to begin 
a catastrophic runaway reaction.  We thus need some assumption
about the rapidity distribution of strangelets produced, and
some assumption about how likely a strangelet of any given 
rapidity is to survive the slowing process and begin a 
runaway reaction.  These assumptions should also apply to any
strangelets produced at RHIC.  They lead to estimates for
``suppression factors'' --- the fraction of those strangelets
produced which survive to begin a runaway reaction --- in the
two cases.  

Here, BJSW and DDH take different approaches.  DDH consider
the worst case hypothesis: a rapidity distribution which 
implies that no strangelet produced on the Moon will survive to 
begin a runaway reaction, while all strangelets produced at RHIC
will.  BJSW propose the rapidity distribution 
\begin{equation}\label{rapidity}
{d \Pi \over d y } = N p y^a e^{-by} \, ,
\end{equation}
where $a,b$ are free parameters, $2p$ is the strangelet production
probability per collision, and $N$ is a normalisation constant.
They model the survival probability as a function of rapidity, 
atomic number and charge as 
\begin{equation}
P ( y , A, Z) = \exp ( - 4.85 ( 1 + {1/3} A^{1/3} )^2 ( \cosh y - 1)
{A/ Z^2 } ) \, . 
\end{equation}
This expression is derived on the assumption that
the strangelet break-up cross-section is independent of energy.  
\end{itemize}

BJSW proceed by calculating the various relative suppression factors for
$a = 1,2,3,4$, and conservatively using the largest of 
these factors (corresponding to $a=4$) in their calculations.  

They conclude that, if iron-iron collisions at RHIC energies are a 
good model for RHIC collisions, the probability of producing a 
dangerous strangelet at RHIC energies can be bounded by
$p_{\rm catastrophe} < 10^{-5}$.   
If only `gold'-iron (or `gold'-`gold') collisions at RHIC energies
are taken as relevant, no bound can be derived.  

BJSW also calculate bounds using collisions at the lower energies of the AGS
experiment, which they believe are (slightly) more plausible for
strangelet production.  They obtain 
$p_{\rm catastrophe} < 2 \times 10^{-11}$ if iron-iron is a good
model, $p_{\rm catastrophe} < 2 \times 10^{-6}$ if `gold'-iron is a good
model, and no useful bound if only `gold'-`gold' collisions are 
relevant.  

BJSW summarise\cite{bjswtwo} the situation thus: ``If, however, one
insists on recreating exactly the circumstances at RHIC and insists
on the worst case rapidity distribution, then lunar limits are not
applicable.''  They add, in the penultimate paragraph of 
Ref. \cite{bjswtwo}, ``The
rapidity distribution necessary to wipe out lunar limits is
bizarre.''  

But no assumption about the rapidity
distribution is necessary to wipe out the lunar limits.
They would not apply if 
there were a significant difference between the dangerous strangelet production
rates for collisions involving iron and those involving gold (or 
`gold').  If only `gold'-`gold' collisions contribute significantly
to the rate of strangelets which are produced and survive slowing,
then, even granting BJSW's rapidity distribution, none of their 
bounds are valid.   

Other questionable assumptions in BJSW's analysis
are whether the precise forms of their rapidity distribution or
survival probability function are correct, and
whether, if so, $a$ must necessarily lie in the 
range $1 \leq a \leq 4$.  To give a convincing argument that 
$p_{\rm catastrophe}$ is less than $< 2 \times 10^{-6}$ (let alone
$2 \times 10^{-11}$), one would need to argue that we can be 
all but certain that each of these assumptions is true.  
BJSW offer no such argument.  

\section{Risk bounds derived from supernova observations} 

Dar, De Rujula and Heinz's analysis\cite{ddh} of the killer strangelet
scenario starts by noting that the bounds based on lunar survival 
may contain a potential loophole.  They suggest
that the rapidity distribution eq. (\ref{rapidity}) might 
be replaced by a distribution which approximates a delta function:
\begin{equation}\label{rapidityddh}
{d \Pi \over d y } =  p \delta ( y - Y /2)  \, ,
\end{equation}
$Y$ being the total rapidity interval.  If so, no strangelets produced
on the Moon by cosmic ray collisions would survive slowing and cause
a catastrophic runaway reaction, but strangelets produced at RHIC
could.  
BJSW, while finding this hypothesis impossible to justify 
theoretically, review DDH's analysis and reconsider their proposed
bounds.  

As noted above, one need not adopt DDH's rapidity
distribution (\ref{rapidityddh}) in order to challenge the 
lunar bounds, which are open to question for a variety of reasons. 
So their independent analysis, which claims ``to derive a fool-proof
and stringent limit on the potential danger of the BNL experiments'', 
needs and deserves careful examination. 

DDH's idea is the following.  High energy collisions between heavy
nuclei cosmic rays occur in space.  If these collisions are capable
of producing killer strangelets, some of the strangelets produced in 
the past will have 
survived slowing interactions with ambient hydrogen and
eventually will have been swept up into protostellar material.
If killer strangelets are capable of destroying the Earth 
catastrophically, they will be equally capable of causing 
catastrophic supernova-like explosions, which we should see.  
Even if killer strangelets are capable of causing only a very
slow destruction of the Earth, over tens or hundreds of millions
of years, they would add significantly to the luminosity of 
stars.  So, the observed rate of supernovae, and the observed
luminosities, should allow us to derive bounds on the probability
of strangelet-induced catastrophe at RHIC.  

Though DDH claim their limit is fool-proof, it actually
relies on some important assumptions.  First, they
assume that killer strangelets produced with relative speed
$v < v_{\rm crit} = 0.1 c$ with respect to ambient 
hydrogen will survive slowing collisions until it 
reaches the local mean velocity.  Second, they assume
that killer strangelets will then survive (growing as they
occasionally hit and absorb hydrogen nuclei) during the long 
time interval before they form part of a proto-star.  

The first of these assumptions is justified by assuming that
the strangelet will have binding energy per nucleon similar
to that of a typical nucleus, and will behave similarly in
collisions with other nuclei.  At first sight, neither claim seems
beyond reasonable doubt.  Yet the assumption is crucial in deriving
DDH's bounds, since the rate of dangerous strangelet production
depends strongly on the value assumed for $v_{\rm crit}$ (cf. 
equation (3) of Ref. \cite{ddh}).  

This seems to be the weakest
point in DDH's argument. However, their second assumption 
could also, in principle, be invalid, for at least three reasons.  

First, killer strangelets could be
metastable rather than stable.  If their lifetime were 
somewhere between, say, $10^{-6}$ seconds and $10^6$ years,
strangelets produced at RHIC would still cause catastrophes,
while almost no strangelets produced in space would survive
long enough to produce observable effects.\cite{aknote}

Second, it could be that the spectrum of stable strangelets is such
that the strangelets produced in collisions
tend to interact with hydrogen nuclei so as to produce 
harmless end-products, although they tend to produce larger
negatively charged strangelets when interacting with other
nuclei.  Again, if so, strangelets produced in space would
not survive to cause catastrophes, while strangelets produced
on Earth would.  

Third, it could be that, in the process of absorbing ambient hydrogen
nuclei and producing decay products, an initially slowly moving 
strangelet would tend to be re-accelerated  with respect to the 
ambient rest frame by randomly directed impulses from the decays.  
If so, it might tend to reach speeds greater than the 
relevant $v_{\rm crit}$, and then break up harmlessly in 
subsequent collisions with further hydrogen nuclei, 
long before forming part of a proto-star.  
It is not obvious that, if this was the typical fate of a 
strangelet produced in space, a similar fate should necessarily 
be expected for strangelets produced at RHIC: the respective
environments are very different.  

\section{Possible threat from a positively charged strangelet?} 

Empirical evidence aside, one of the theoretical arguments against
the ``killer strangelet'' catastrophe scenario is that the existence of
negatively charged strangelets would be theoretically surprising.  
Such a strangelet would have to contain a proportion greater
than $1/3$ of $s$ quarks, which seems unlikely, since the higher
mass of the $s$ quark suggests their proportion should be lower.

For this reason, it is generally believed that, if any strangelets
were produced in RHIC or other collider experiments, they would
be positively charged.  
Neither BJSW nor DDH seriously consider the possibility
of producing positively charged strangelets, since they state
that positively charged strangelets would pose no threat 
whatsoever.  

The reason for this is that a positively charged strangelet, embedded
in solid matter --- for instance, the wall of a collider --- 
would simply behave like an exotic nucleus.  Coulomb repulsion
would prevent it from approaching other nuclei and fusing with
them, and so there would be no runaway reaction.  

This argument does not, however, apply in all environments.  In
particular, if a positively charged strangelet reached the interior of
the Sun, it seems quite possible that it would initiate a catastrophic
runaway reaction, in the same way that a negatively charged strangelet
would.  So long as the Coulomb repulsion is overwhelmed by
gravitational pressure (as it certainly would be initially) 
the strangelet charge would be essentially irrelevant.  

The probability of a strangelet produced on Earth reaching the Sun
without active human intervention, before the Earth is consumed
by the Sun (by when it presumably would not matter), is presumably
negligible.  However, the possibility of human intervention needs to 
be considered.   We do send spacecraft towards
and even into the Sun, and objects in the vicinity of particle
physics colliders --- for instance, reused detector 
components --- are considerably more 
likely than most to be included in these probes.  
Laboratory material potentially 
contaminated by positively charged strangelets would thus 
pose a potential danger which, though small, would need careful 
handling.\footnote{Another possible concern might be that 
a malicious entity could, at some point in the future, obtain 
material contaminated by positive strangelets and then deliberately 
threaten to cause catastrophe.  
Presumably, though, any group with the technological resources to 
send a spacecraft to the Sun, or to cause a fusion reaction 
on Earth, could credibly threaten catastrophe 
by less exotic means, if so inclined.} 

It should be stressed that BJSW give two theoretical arguments
which suggest that the production of positively charged strangelets at
RHIC, while less implausible, is still unlikely.
These are the fact that theory seems to
weigh against the possibility of stable strange matter of any type,
and the fact that theory argues against the possibility of
producing any form of strangelet in heavy ion collisions. 
BJSW find the second of these arguments particularly compelling. 
If either conclusion is correct, then, of course, there is
no cause at all for concern.  Still, theorists'
opinions do seem rather unsettled on this question. 
As recently as July 1999, the possibility of producing a (presumably
positively charged) strangelet at RHIC was described 
as ``a speculative but quite respectable 
possibility''.\cite{wilczek} 

Interestingly, DDH's empirical arguments against the possibility 
of producing negatively charged strangelets at RHIC 
apply equally well to positively charged strangelets ---
though, of course, the weaknesses noted above remain. 
BJSW's empirical arguments do not apply, since a positively
charged strangelet created on the Moon would cause no 
catastrophe.  Weaker bounds using similar reasoning might,
however, be derived given that asteroids are also potentially
vulnerable to strangelet production via cosmic ray impact, and
that the infall rate of asteroid matter into the Sun 
has been estimated.\cite{asteroidinflux}  

It is worth noting here that stronger bounds could be
derived by assuming that our solar system is not very atypical 
and then combining BJSW's and DDH's lines of argument.  Suppose that
a significant proportion of stars are orbited by asteroids, satellites or
planetoids unprotected by atmosphere, that those bodies tend
to have similar chemical compositions to those of our solar system,
and that the stellar infall rate of these bodies is similar to that 
of our solar system.  Quantifying these assumptions, we could estimate 
the rate at which strangelets, created by cosmic ray impact on
asteroids, were introduced into stars, causing supernovae.  
The bounds derived would apply whether the strangelets created are positively
or negatively charged, assuming that the infall rate of asteroids
converted into strange matter would be at least as great as that of 
unaffected asteroids.  
Unfortunately, though it seems very likely that our solar system
is indeed not atypical, we have no firm evidence.  Moreover, 
the potential flaws in DDH's arguments, noted above, also 
apply to this line of reasoning.  Nonetheless, it would be interesting to 
pursue it further.  

\section{Conclusions} 

To reiterate: this paper is written from the point
of view of a devil's advocate trying to point 
out potential flaws in the empirical arguments
of BJSW and DDH.  
Those arguments rely on hypotheses which seem 
reasonable, but which could be wrong.  To place any confidence
in BJSW's and DDH's quantitative bounds 
on the risk of catastrophe, we need to be confident that the 
probability of their hypotheses being wrong is very low.  
BJSW and DDH have not carefully addressed this question. 

Among the potential flaws in BJSW's lunar survival arguments are
(i) the possibility that the behaviour of iron nuclei may not be a 
reasonable model for that of gold nuclei; (ii) the possibility that 
their best guess at the strangelet rapidity distribution may
be significantly wrong (without necessarily taking the extreme
form suggested by DDH); (iii) the possibility that, although 
their distribution is roughly correct, the parameter $a$ lies
outside the range they consider; (iv) the possibility that their 
strangelet survival probability function may be significantly wrong. 

Among the potential flaws in DDH's arguments based on supernovae
observations are (i) the possibility that the critical speed above
which strangelets are broken up in collisions with hydrogen nuclei
might be substantially lower than their guess; (ii) the possibility
that ``killer strangelets'' might be metastable; (iii) the possibility
that they might tend to produce harmless end-products after
combining with hydrogen nuclei, but not after combining with 
larger nuclei; (iv) the possibility that the process of absorbing
stray hydrogen nuclei and emitting decay products might tend 
to re-accelerate a slowed strangelet in interstellar space, 
leading to its eventual breakup in a collision with a hydrogen nucleus. 

It would be reassuring if all of these possibilities could 
be convincingly rebutted.  

\section{Acknowledgements} 

Friends and colleagues too numerous to list here have provided invaluable
help.  I am very grateful to them all.  This work was supported 
by the Royal Society and PPARC.



\begin{thebibliography}{99}
\bibitem{aknote}
This point was made in private correspondence to BJSW, and 
subsequently discussed in the second version of their report;
see reference 5 of Ref. \cite{bjswtwo}.
\end{thebibliography}
\end{document}